\documentclass[12pt]{article}
\usepackage{geometry}                % See geometry.pdf to learn the layout options. There are lots.
\usepackage{graphicx}
\usepackage{amssymb}
\usepackage{epstopdf}
\def\szero{s^{(0)}_{ab} }

\def\hk{ H^{(k)}}

\def\call{{\cal L}}
\def\callk{{\cal L}^{(k)}}
\def\calg{{\cal G}}

\def\b2hat{ {\hat b}_2 }

\def\szero{ s^{(0)}_{ab} }

\begin{document}

\begin{titlepage}
\vfill
\begin{flushright}
\today
\end{flushright}

\vfill
%\vskip 1.0cm
\begin{center}
\baselineskip=16pt
{\Large\bf Smarr Formula and an Extended First Law}
\vskip 0.15in
{\Large\bf for Lovelock Gravity}
\vskip 0.5cm
{\large {\sl }}
\vskip 10.mm
{\bf David Kastor${}^{a}$, Sourya Ray${}^b$ and Jennie Traschen${}^a$} \\

\vskip 1cm
{
	${}^a$ Department of Physics, University of Massachusetts, Amherst, MA 01003\\	
     	${}^b$ Centro de Estudios Cient\'{\i}ficos (CECS), Casilla 1469, Valdivia, Chile \\
	\texttt{kastor@physics.umass.edu, ray@cecs.cl, traschen@physics.umass.edu}

     }
\vspace{6pt}
\end{center}
\vskip 0.2in
\par
\begin{center}
{\bf Abstract}
 \end{center}
\begin{quote}
We study properties of static, asymptotically AdS black holes in Lovelock gravity. Our main result is a Smarr formula 
that gives the mass in terms of geometrical quantities
together with the parameters of the Lovelock theory.  
%The derivation is complicated by the infinite amount of  stress-energy 
%present in the region between the horizon and asymptotic AdS region.  
%In order to avoid divergences, we work with geometrical
%quantities that appear in the constraint equations, and to work with identities that follow from
%the constraints, so that infinities do not enter at intermediate steps.
As in Einstein gravity, the Smarr formula follows from applying the first law to an infinitesimal change in the overall length scale.
However, because the Lovelock couplings are dimensionful, we must first prove an extension of the 
 first law that includes their variations.  Key ingredients in this construction are the Killing-Lovelock potentials 
 associated with each of the higher curvature Lovelock interactions.  Geometric expressions are obtained for the new thermodynamic potentials conjugate to variation of the Lovelock couplings.
  \vfill
% \hrule width 5.cm
\vskip 2.mm
\end{quote}
\end{titlepage}

%\maketitle
%\section{}
%\subsection{}

\section{Introduction}
Lovelock theories  \cite{Lovelock:1971yv} are an intriguing subset of higher curvature gravity theories.  While the field equations for a general higher curvature theory 
involve second derivatives of the Riemann tensor, Lovelock theories share the property of Einstein gravity that no derivatives of the curvature tensor, and hence only second derivatives of the metric tensor, arise.  It follows that  Lovelock gravities share a number of additional nice properties with Einstein gravity that are not enjoyed by other more general higher curvature  theories.  Most prominently, Lovelock gravities can have stable, ghost free, constant curvature vacua \cite{Zwiebach:1985uq}\cite{Zumino:1985dp} and hence are suitable starting points for quantization\footnote{Not all constant curvature vacua of Lovelock gravity theories are stable and ghost free.  A recent analysis \cite{Charmousis:2008ce} of the Lovelock gravity theory including up to curvature squared terms found that in the region of parameter space supporting two constant curvature vacua, typically  one was stable, and one was unstable.  See also reference \cite{Canfora:2008iu} for a related discussion  including the Lovelock curvature cubed terms in $D=7$ with three dimensions compactified.}.
For the purposes of this paper it will be particularly useful that the second order character of the Lovelock field equations  leads to a reasonably well behaved Hamiltonian formulation of the theory \cite{lovelock-hamiltonian}.

Lovelock gravity theories have been studied by many authors over a wide span of years.  In particular, exact, static black hole solutions were found beginning with the work of  \cite{Boulware:1985wk}\cite{Wheeler:1985nh}\cite{Wheeler:1985qd} (see also the recent reviews \cite{Charmousis:2008kc}\cite{Garraffo:2008hu}).  An expression for the entropy of Lovelock black holes was obtained in \cite{Jacobson:1993xs} by means of integrating a first law that was derived using the Hamiltonian perturbation theory techniques of \cite{Sudarsky:1992ty} (see also reference \cite{Cai:2003kt} for the entropy of asymptotically AdS black holes with flat and negatively curved horizons).  The entropy formula includes new contributions coming from the higher curvature terms in the action and is in agreement with the entropy formula obtained later via more general methods in \cite{Wald:1993nt}\cite{Iyer:1994ys}. 
There has also been considerable recent interest in Lovelock gravities in the context of the AdS/CFT  correspondence \cite{Brigante:2007nu,Brigante:2008gz,Buchel:2009tt,Hofman:2009ug,Shu:2009ax,deBoer:2009pn,Camanho:2009vw,Buchel:2009sk,deBoer:2009gx,Camanho:2009hu,Ge:2010aa}, while stability of static Lovelock black holes under perturbations has been addressed in \cite{Dotti:2004sh,Dotti:2005sq,Gleiser:2005ra,Beroiz:2007gp,Takahashi:2009dz,Takahashi:2009xh}.

This paper extends to Lovelock gravities a line of inquiry that was initiated in references \cite{Kastor:2008xb,Kastor:2009wy}.  In 
\cite{Kastor:2009wy} we studied certain thermodynamic properties of black holes in Einstein gravity with a non-zero cosmological constant $\Lambda$.
Geometrical techniques were used to derive a Smarr formula and also a related extension of the first law that includes variations $\delta\Lambda$ in the cosmological constant.
The resulting Smarr formula has the form
%%
%\begin{equation}\label{basicsmarr}
%(D-3)M = (D-2)TS - 2 {\Theta\over 8\pi G}\Lambda 
%\end{equation}
%%
%%
\begin{equation}\label{basicsmarr}
(D-3) GM = (D-2){\kappa A\over 8\pi} - 2 {\Theta\Lambda\over 8\pi} 
\end{equation}
where $M$ is the ADM mass of the black hole, $\kappa$ and $A$ are the surface gravity and area of the horizon, and 
%as usual the temperature $T$ is given in terms of the surface gravity by $T=\kappa/2\pi$ and the entropy is given in terms of the horizon area by $S=A/4G$, and now
$\Theta$ represents a new potential that is thermodynamically conjugate to $\Lambda$.
For $\Lambda=0$ equation (\ref{basicsmarr}) reduces to the well known result \cite{Smarr:1972kt} for static, asymptotically flat black holes.  The extension to $\Lambda\neq 0$ had been previously established based on analysis of the explicit Schwarzschild-(A)dS solutions  \cite{Caldarelli:1999xj,Wang:2006eb,Sekiwa:2006qj,Wang:2006bn,Larranaga Rubio:2007jz,Cardoso:2008gm,Urano:2009xn}.  However, no general derivation was given, and although explicit expressions for the potential $\Theta$ were obtained in terms of $M$ and $\Lambda$, its geometric significance could not be addressed by these means.

There are two ways to derive the Smarr formula for $\Lambda=0$, both of which were generalized to $\Lambda\neq 0$ in \cite{Kastor:2009wy}.  Both of these derivations may also be extended to Lovelock gravity theories.  Although we will only present the second method here for the Lovelock case, it will be worthwhile to briefly discuss the first method as well.
The first method makes use of Komar integral relations \cite{Komar:1958wp}, which are Gauss' law or Stokes' theorem type statements that hold on a spatial slice in a spacetime with a Killing vector.  For a static black hole with $\Lambda=0$, the Komar integral relation for the time translation Killing vector yields an equality between a boundary integral at infinity proportional to $M$ and one at the black hole horizon proportional to $\kappa A$.  This equality is the Smarr formula (\ref{basicsmarr}) with 
$\Lambda=0$.  However, this approach does not extend directly to $\Lambda\neq 0$.  The cosmological constant acts as a source term in the Komar integral relation, present everywhere on the spatial slice.  Considering AdS black holes, the boundary integral at infinity is no longer equal to the horizon boundary integral, because of this source. The boundary integral at infinity, in fact, diverges.  It can be regularized, by integrating over a sphere at large but not infinite radius, and renormalized, via a background subtraction, to yield a finite mass \cite{Magnon:1985sc}.  However, this procedure does not yield a Smarr formula.

An improved Komar integral relation for the $\Lambda\neq 0$ case was introduced in \cite{Kastor:2008xb} and applied in \cite{Kastor:2009wy} to establish equation (\ref{basicsmarr}).  In the improved relation, the volume source term proportional to $\Lambda$ is written as a total divergence and converted to a boundary integral.  The new boundary integrand is proportional to the antisymmetric Killing potential $\beta^{ab}$, related to the Killing vector 
$\xi^a$ through
\begin{equation}\label{kp}
\xi^b=\nabla_a \beta^{ab}.
\end{equation}
The combination of this new boundary integral with the original Komar integral at infinity now yields a finite result, eliminating the need for regularization and renormalization.  The improved Komar integral relation then yields the Smarr formula (\ref{basicsmarr}), including an expression for the potential $\Theta$ in terms of the difference between integrals of the Killing potential at the horizon and infinity.

The second method of establishing the Smarr formula for $\Lambda=0$ is via the first law
%$\delta M = T\delta S$, or equivalently
$\delta (G M) = \kappa \delta A/8\pi $.  
In order for the first law to be satisfied by an overall change in length scale $L$, in which the quantities $GM$ and $A$ scale as $L^{D-3}$ and $L^{D-2}$ respectively, the mass and area must be related as in (\ref{basicsmarr}).  As noted in \cite{Gibbons:2004ai}, in order to apply this method with $\Lambda\neq 0$ one needs a version of the first law in which the cosmological constant $\Lambda$, which scales as $L^{-2}$, is also allowed to vary.  The derivation of such an extended first law in \cite{Kastor:2009wy} also makes use of the Killing potential $\beta^{ab}$ and ultimately yields the same expression for the potential $\Theta$ in equation (\ref{basicsmarr}) as using Komar integrals.

The construction of improved Komar integral relations given in \cite{Kastor:2008xb} applies to all Lovelock gravity theories,  the addition of the cosmological constant term to Einstein gravity being the simplest case in which it is necessary.  The key ingredient in the construction is a sequence of Killing-Lovelock potentials $\beta^{(k)ab}$  (defined below) that are in one-to-one correspondence with the higher curvature terms $\call^{(k)}$ (also given below) in the Lagrangian of a Lovelock gravity theory.
We will follow the first law method described above to derive the Smarr formula for static, asymptotically AdS Lovelock black holes.  
The required extension of the first law includes variations $\delta b_k$ of each of the dimensionful coupling constants $b_k$ of the Lovelock theory.   It has the general form
\begin{equation}\label{firstlawform}
\delta (G M) = {\kappa\delta\hat A \over 8\pi}- {1\over 16\pi}\sum_k \Psi^{(k)}\delta b_k .
\end{equation}
where the quantity $\hat A$ defined in equation (\ref{ak}) below is related to the black hole entropy by $S= \hat A/4G$, and
the quantities $\Psi^{(k)}$ are new thermodynamic potentials conjugate to the individual Lovelock couplings $b_k$.  The central result of the paper is the expression
(\ref{psik}) that we derive for the $\Psi^{(k)}$.   We find that there are three contributions to this result.  The first, which we write as $\Theta^{(k)}$, is given in terms of boundary integrals of the Killing-Lovelock potentials.  It can be thought of as coming from the variation of the effective stress-energy introduced by the higher curvature Lovelock interactions.  The second and third contributions come from the explicit dependence of the mass and entropy on the Lovelock couplings.  In the simple case of Einstein gravity with a non-zero cosmological constant $\Lambda$  studied in  \cite{Kastor:2009wy}, only the first of these contributions arose, because the mass and entropy have no explicit dependence on $\Lambda$.
 
%The final result for the extended first law and the Smarr formula includes finite potentials $\Theta^{(k)}$ that are conjugate to each of the Lovelock coupling constants and are determined by boundary integrals of the corresponding Killing-Lovelock potentials.  We will see explicitly how the Killing-Lovelock potential terms cancel  divergences at infinity that come from varying the Lovelock couplings.  

Given the extended first law, it is then straightforward to derive the Smarr formula via the scaling argument discussed above.
The quantity $\hat A$ scales in the same way as the horizon area, while the Lovelock couplings scale as $L^{2(k-1)}$, giving the result
\begin{equation}\label{lovelocksmarr}
(D-3)M = (D-2) TS -\sum_k 2(k-1){\Psi^{(k)} b_k\over 16\pi G}
\end{equation}
%
%%
%\begin{equation}
%(D-3)GM = (D-2) {\kappa\hat A\over 8\pi} -\sum_k 2(k-1){\Psi^{(k)} b_k\over 16\pi}
%\end{equation}
%%
where the relation between the temperature and surface gravity $T=\kappa/2\pi$ has now been used.
The  Smarr formula gives the mass of a static, asymptotically AdS black hole in terms of certain thermodynamic properties,  the surface gravity, entropy, and the potentials  $\Psi^{(k)}$.  It becomes particularly useful when such solutions are not known explicitly, as is the case in higher order Lovelock theories \cite{Wheeler:1985qd}.   In the concluding section we will indicate how the Smarr formula may be used in analyzing the Hawking-Page 
\cite{Hawking:1982dh} phase transition  in general Lovelock gravity theories.

The remainder of the paper is organized as follows. Section (\ref{seclg}) summarizes the basic formalism of Lovelock
gravity including, in particular, the Killing-Lovelock potentials and the Hamiltonian formulation of the theory. 
In section (\ref{secfirst}) we use Hamiltonian perturbation theory techniques to prove the extended first law for Lovelock gravities, including variations
in the Lovelock couplings.   The final result for the Smarr formula then follows from the scaling argument sketched above.
Section (\ref{secfree}) contains concluding remarks including some directions for future research.

\section{Lovelock Gravity}\label{seclg}
Lovelock gravities \cite{Lovelock:1971yv} are the unique higher curvature gravity theories with field equations that do not involve derivatives of the Riemann curvature tensor.  In the subsections below, we will present the parts of the formalism of Lovelock gravity that are needed for  our derivations of the extended first law and Smarr formula.

\subsection{Lagrangian and equations of motion}\label{basics}

The Lagrangian  of a Lovelock gravity theory  is given by ${\cal L}=\sum_{k\ge 0} c_k{\cal L}^{(k)}$,  where the $c_k$ are real coefficients specifying the theory and the $\callk$ are particular scalar contractions of $k$ powers of the Riemann tensor given by 
\begin{equation}\label{lovelagran}
\call^{(k)} ={1\over 2^k } \delta ^{a_1 b_1...a_k b_k } _{c_1 d_1 ....c_k d_k }
 R_{a_1 b_1}{}^{c_1 d_1 }\dots  R_{a_k b_k}{}^{c_k d_k }.
\end{equation}
Here  the $\delta$ symbol is the totally anti-symmetrized product
of Kronecker deltas 
$\delta ^{a_1 ... a_n } _{c_1 ... c_n }=n! \delta ^{[a_1}_{c_1} \dots\delta ^{a_n]}_ {c_n }$, which is normalized so that it takes nonzero values $\pm 1$.  By definition, the zeroth  order term in the curvature is given by  $\call^{(0)}=1$, while the linear term is normalized so that $\call^{(1)} = R$.   The term quadratic in the curvature, known as the Gauss-Bonnet term, is given by $\call^{(2)} = R_{ab}{}^{cd}R_{cd}{}^{ab} - 4 R_a{}^bR_b{}^a +R^2$.
%%
%\begin{equation}\label{gbterm}
%\call^{(2)} = R_{ab}{}^{cd}R_{cd}{}^{ab} - 4 R_a{}^bR_b{}^a +R^2
%\end{equation}
%%

A given term $\callk$ in the Lovelock Lagrangian can only contribute to the dynamics in dimensions $D>2k$.  For $D<2k$, $\callk$ is easily seen to vanish identically, while for $D=2k$ its integral is proportional to the topologically invariant Euler character of the manifold.   It follows that the variation of  $\callk$ in $D=2k$ is a total derivative and does not contribute to the equations of motion.  This is the case {\it e.g.} for the Einstein term $\call^{(1)} = R$ in $D=2$ and for the Gauss-Bonnet term $\call^{(2)}$ in $D=4$.  The Gauss-Bonnet term only contributes  to the dynamics of theories for $D\ge 5$.
More generally, the Lovelock lagrangian in $D$ spacetime dimensions can be taken to be the truncated sum
\begin{equation}
{\cal L}=\sum_{k= 0}^{\hat k}c_k{\cal L}^{(k)}
\end{equation}
where $\hat k = [(D-1)/2]$.  
The field equations for Lovelock gravity have the form $\calg^a{}_b=0$ where 
$\calg^a{}_b = \sum_k c_k{\cal G}^{(k)a}{}_b$, with 
the individual Einstein-like tensors ${\cal G}^{(k)a}{}_{ b}$ given by
\begin{equation}\label{lovelockmotion}
\mathcal{G}^{(k)a}{}_b = -{1\over 2^{(k+1)}}
\delta^{ac_1d_1...c_kd_k}_{be_1f_1...e_kf_k}\,R_{c_1d_1}{}^{e_1f_1}\dots R_{c_kd_k}{}^{e_kf_k}.
\end{equation}
Because the field equations do not contain any derivatives of the curvature tensor, 
we see explicitly that the equations of motion do not involve any derivatives of the metric higher than second order.
Since the terms $\callk$ are diffeomorphism invariant, each of the Einstein-like tensors ${\cal G}^{(k)a}{} _b$ satisfies  
 $\nabla _a {\cal G}^{(k)a}{} _b  =0$.

 \subsection{Including a factor of Newton's constant}
In the case of pure gravity,  an overall rescaling of  the gravitational action does not 
change the equations of motion.  When gravity is coupled to matter a factor of $1/16\pi G$ is typically written in front
of the gravitational action, which defines the strength of the coupling of gravity to matter.
Even though we are studying vacuum spacetimes in this paper, it will be useful later on
to work with rescaled Lovelock parameters $b_k$ defined by
\begin{equation}\label{bkdef}
c_k ={b_k \over16\pi  G}
\end{equation}
so that an overall factor of $1/16\pi G$ will multiply the gravitational action.
The parameters $b_k$ then have dimensions $L^{2(k-1)}$, with $b_1$ being dimensionless.
 In general, the total Lagrangian is a sum of the Lovelock Lagrangian plus
a contribution from matter fields, and
then the field equations are ${\cal G}^a_{\ b}=\sum _k b_k  {\cal G}^{(k)}  _{ab} =8\pi G T  _{ab}$.
Of course the distinction between the $c_k$ and the $b_k$ is hidden when one works
in units with $G=1$, which gives mass the same units as length, but keeping factors of $G$ explicit
will be important for identifying physical quantities and their scaling dimensions.

 \subsection{Black hole entropy}
 We will also need the formula for black hole entropy in Lovelock gravity, which was obtained in reference \cite{Jacobson:1993xs} by integrating the first law for Lovelock black holes (with fixed Lovelock couplings).  First define the quantities 
 \begin{equation}\label{ak}
 \hat A = \sum_k b_k A^{(k)} ,\qquad A^{(k)} = k\int_H da \call^{(k-1)}
 \end{equation}
 %
% %
% \begin{eqnarray}\label{ak}
% A^{(k)} &=& k\int_H da \call^{(k-1)}\\
% \hat A &=& \sum_k A^{(k)}
% \end{eqnarray}
% %
 where the integral is taken over the intersection of the black hole horizon with a spacelike slice and the Lovelock term $\call^{(k-1)}$ is evaluated with respect to the induced metric on this cross-section of the horizon.   The black hole entropy is then given by
 \begin{equation}\label{entropy}
 S = {1\over 4G}\hat A
 %S= \sum_k 4\pi c_k A^{(k)}
 \end{equation}
 Note that $\call^{(0)}=1$ and therefore $A^{(1)}$ is simply the horizon area, reproducing the familiar result of Einstein gravity, while the Gauss-Bonnet term in the original Lovelock action will make a contribution to the entropy proportional to the integral of the scalar curvature over the horizon.  The fact that the entropy has explicit dependence on the Lovelock couplings will be important below when we consider the effect of varying these couplings.
 
% \begin{equation}\label{entropy}
% S = \sum_{k=1}^{\hat k} 4\pi k c_k\int d^{D-2}x \sqrt{\gamma}\call^{(k-1)}(\gamma_{ab})
% \end{equation}
% %
% where the integral is taken over the intersection of the black hole horizon with a spacelike slice and $\gamma_{ab}$ is the induced metric on this cross-section of the horizon.  Recalling that $\call^{(0)}=1$, one sees that the contribution of the Einstein term in the action is indeed proportional to the area of the event horizon, while the contribution of the Gauss-Bonnet term in the action is proportional to the integral of $\call^{(1)}=R$.

 \subsection{Killing-Lovelock potentials}\label{killingpotential}
 
The Killing-Lovelock potentials \cite{Kastor:2008xb} which play an important role in our construction arise in the following way.
Stokes theorem implies that  a divergence free vector field may be written, at least locally, as the divergence of an anti-symmetric two index tensor.
Killing vectors are divergenceless and hence a Killing vector $\xi^a$ can be expressed in terms of an anti-symmetric Killing potential as 
$\xi^b = \nabla_a\beta^{ab}$.  The Killing potential $\beta^{ab}$ is not unique, but can be shifted by the addition of a divergenceless tensor.

Further, each Einstein-like tensors $ {\cal G}^{(k)a}{} _b$  may be contracted with the Killing vector $\xi^a$ to give the divergenceless currents
 ${\cal J}^{(k)a} = {\cal G}^{(k)a}{} _b\xi^b$.  The Killing-Lovelock potentials $\beta^{(k)ab}$ are then defined to be solutions of
\begin{equation}\label{potdef}
{\cal J}^{(k)b}  = - {1\over 2} \nabla_a \beta^{(k)ab}
\end{equation}
where the coefficient $-1/2$ on the right hand side is introduced for later convenience.   The Killing-Lovelock potentials are defined only up to addition of an arbitrary divergenceless tensor.  From equation (\ref{lovelockmotion}), we see that ${\cal G}^{(0)a}{} _b=-(1/2)\delta^a_b$ and hence $ \beta^{(0)ab}$ is simply the Killing potential.  The Killing potential played an important role in the derivation of the Smarr formula with $\Lambda\neq 0$ and the first law with varying $\Lambda$ in reference \cite{Kastor:2009wy}.  The Killing potentials with $k>0$ will play similar roles in our construction below.

\subsection{Constant curvature vacua}\label{vacua}

A Lovelock gravity theory in $D$ spacetime dimensions, specified by a set of coefficients $c_0,\dots,c_{\hat k}$, can have between $0$ and $\hat k$ distinct constant curvature vacua. These vacua determine the possible asymptotic behaviors for black hole solutions of Lovelock gravity.
The possible vacuum curvatures may be found by rewriting the overall Einstein-like tensor $\calg^a{}_b$ in the form
\begin{equation}\label{alteom}
\calg^a{}_b = \alpha_0\delta^{ac_1d_1\dots c_{\hat k}d_{\hat k}}_{fe_1f_1\dots e_{\hat k}f_{\hat k}}
(R_{c_1d_1}{}^{e_1f_1}-\alpha_1\delta^{e_1f_1}_{c_1d_1})\dots
(R_{c_{\hat k}d_{\hat k}}{}^{e_{\hat k}f_{\hat k}}-\alpha_{\hat k}\delta^{e_{\hat k}f_{\hat k}}_{c_{\hat k}d_{\hat k}}).
\end{equation}
Through use of the identity 
\begin{equation}
\delta^{a_1\dots a_p}_{b_1\dots b_p}\delta^{b_{p-1}b_{p}}_{a_{p-1}a_{p}}
=2(D-p+1)(D-p+2)\delta^{a_1\dots a_{p-2}}_{b_1\dots b_{p-2}}.
\end{equation}
one can see that this is equivalent to the original form given in section (\ref{basics}).
The real-valued coefficients $b_k$ in the Lovelock Lagrangian are given by sums of products of the new coefficients $\alpha_k$ in (\ref{alteom}) (the explicit relations are given in \cite{Crisostomo:2000bb}).
Inverting this relation to get the $\alpha_k$'s in terms of the $b_k$'s would require solving a system of algebraic equations that are of order $\hat k$ in their variables.  Hence some, or all, of  the coefficients $\alpha_k$ may be complex.
If some particular coefficient $\alpha_p$ is real, it follows that a spacetime with constant curvature given by 
$R_{ab}{}^{cd} = \alpha_p\delta_{ab}^{cd}$ solves the equations of motion $\calg^a{}_b=0$.
Our primary interest will be in asymptotically AdS black holes, and we will therefore assume that we are working with a Lovelock gravity theory having at least one 
$\alpha_k$ real and negative.

\subsection{Hamiltonian formulation}

The Hamiltonian formulation of Lovelock gravity given in \cite{lovelock-hamiltonian} was used in \cite{Jacobson:1993xs} to derive a first law and the expression 
(\ref{entropy}) for the gravitational entropy in Lovelock theories.  We will make similar use of the Hamiltonian formalism in combination with the Killing potentials to derive an extended first law in which the Lovelock coupling constants are allowed to vary.
As usual, the spacetime metric is split according to
 \begin{equation}\label{metricsplit}
g_{ab}=-n_a n_b +s_{ab}, \qquad n_a n^a =-1, \qquad s_a{}^b n_b =0 .
\end{equation}
where $n^a$ is the unit timelike normal to a spatial slice $\Sigma$ and $s_{ab}$ is the induced metric on $\Sigma$.
The canonical variables are the spatial metric  $s_{ab}$ and its conjugate momentum\footnote{Although it is not relevant for our discussion (or the derivation of the first law with fixed Lovelock coupling constants in \cite{Jacobson:1993xs}), it should be noted that the Hamiltonian formulation of Lovelock gravity is generally ill-defined  \cite{lovelock-hamiltonian} in the sense that the ``velocities'' are multi-valued functions of the momenta.  The exception to this is when all the parameters of the theory are such that all the $\alpha$'s in equation (\ref{alteom}) are equal and there is a unique constant curvature vacuum.}  
$\pi ^{ab}$.
We consider Hamiltonian evolution along a vector field $\xi^a$, which can be written in projected form as 
$\xi ^a = Fn^a +F ^a$ with $F^a =s^a{}_b\xi^b$.   The 
Hamiltonian is then given by ${\cal H} =\int \sqrt{s} H$ where
$H=FH_{\bot}+F^aH_a$ where as in Einstein gravity  the quantities $H_\bot$ and $H_a$ are proportional to the 
components of the Einstein-like tensor $\calg_{ab}$ with one index contracted with the normal $n^a$ and the other index projected normal to or parallel to the surface respectively,
\begin{equation}
H_\bot = -2\calg_{ab}n^an^b ,\qquad H_a = -2s_a{}^bn^c \calg_{bc}.
\end{equation}
Also as in Einstein gravity, the lapse $F$ and shift $F^a $ are Lagrange multipliers.
 One can further decompose the Hamiltonian order by order in the curvature as
 $ H=\sum_k b_k  H^{(k)}$
with $H^{(k)}=FH^{(k)}_{\bot} + F^aH^{(k)}_a$, $H^{(k)}_{\bot}=-2\calg^{(k)}_{ab}n^an^b$ and $H^{(k)}_a = -2s_a{}^bn^c \calg^{(k)}_{bc}$.
Adding in the stress-energy tensor for later reference, the Hamiltonian and momentum constraint equations are given by
$H_{\bot} =-16\pi G \rho$ and $H_a =-16\pi G J_a $, where
$\rho= n^a n^b T_{ab} $ and $J_a = s_a{}^b n^c T_{bc}$.

For a stationary black hole of Lovelock gravity, one can  consider Hamiltonian evolution of the initial data with respect to the horizon generating Killing field $\xi^a$, with corresponding lapse $F$ and shift vector $F^a$.
For this choice of vector field the individual terms $H^{(k)}$ in the decomposition of the Hamiltonian $H$ can each be re-written as a total divergence
in terms of  the Killing potentials $\beta^{(k)ab}$ introduced in section (\ref{killingpotential}).  One has
\begin{equation}\label{kthhamdiv}
H^{(k)}=  F \hk _{\bot} +F^a \hk _a 
= -2 {\cal G}^{(k)d}  _c \xi ^c n_d = D_c ( \beta^{(k)cd} n_d )
\end{equation}  
where $D_a$ is the covariant derivative operator associated with the spatial metric $s_{ab}$.

\section{Lovelock first law with varying couplings}\label{secfirst}

The Smarr formula in Einstein gravity may be obtained from the first law by considering
perturbations that represent an overall change in length scale (see {\it e.g.} references \cite{Gauntlett:1998fz,Townsend:2001rg}). 
Lovelock gravity theories, however, include additional dimensionful couplings, and in order to obtain the Smarr formula in a similar fashion,
one must start with a version of the first law that includes variations in the Lovelock couplings.  The need for such an extended first law already arises for
Einstein gravity with a non-zero cosmological constant \cite{Gibbons:2004ai}. The contribution to the first law from a varying cosmological constant was computed and used to obtain the Smarr formula in reference \cite{Kastor:2009wy}\footnote{One might wonder how the scaling argument in Einstein gravity could have given the correct result without using a first law that includes variations $\delta G$ in Newton's constant, since it is also dimensionful.
We will address this point below.}.  

A first law for Lovelock black holes with fixed coupling constants was derived in \cite{Jacobson:1993xs} using the Hamiltonian perturbation theory methods of
\cite{Sudarsky:1992ty} (see  \cite{Traschen:1984bp} for a more general formulation). Limiting our consideration to perturbations around static black holes, the result of \cite{Jacobson:1993xs} has the familiar form
\begin{equation}\label{lovelockfirst}
 \left. \delta M \right |_{b_k}  = \left. T\delta S   \right |_{b_k} 
\end{equation}
where the temperature is given in terms of the surface gravity $\kappa$ by $T=\kappa/2\pi$, the entropy $S$ is given in equation (\ref{entropy}) and we indicate explicitly that we are considering variations with the Lovelock couplings held fixed.  In this section we will compute the new contributions to (\ref{lovelockfirst}) that arise when the Lovelock couplings are allowed to vary.
%\footnote{ The 
%Komar route was also used for Einstein gravity with cosmological constant in \cite{Kastor:2009wy},
%and indeed could be done for the general Lovelock Lagrangian by using the
%Killing-Lovelock potentials. However, the calculations become cumbersome.}.

\subsection{Hamiltonian perturbation theory}
The Hamiltonian perturbation theory construction works in the following way. Let the metric
be split as in equation (\ref{metricsplit}). The Hamiltonian variables are the metric
on the spacelike slices $s_{ab}$ and its conjugate momentum $\pi ^{ab}$.
 We first  recall the case where the Lovelock coupling constants are held fixed \cite{Jacobson:1993xs} and then show how allowing variations in the couplings modifies the results.
Let the metric $ g^{(0)}_{ab}$ be a solution to  the equations of motion
 with a static Killing vector $\xi^a$. The corresponding Hamiltonian variables
  are $\szero$ and $\pi^{ab}_{(0)}$ and  satisfy $\call_\xi \szero = \call_\xi \pi^{ab}_{(0)} = 0$.
Now consider perturbing the metric such that
$s_{ab} = \szero +\delta s_{ab}$ and $\pi^{ab} = \pi^{ab}_{(0)} +\delta\pi^{ab}$.
The perturbed metric is also assumed to satisfy the equations of motion, but the perturbations are not required to be invariant under the symmetry generated by the Killing vector $\xi^a$.

The variation of $H_\perp$  with respect to these perturbations then takes the form
\begin{equation}
\delta H_{\bot} ={\delta H_{\bot} \over \delta s_{ab} }\cdot \delta s_{ab} +
{\delta H_{\bot} \over \delta \pi^{ab} }\cdot \delta \pi^{ab} 
\end{equation}
and similarly for the variation  $\delta H_a$.  Here  $\cdot$  indicates that the quantity to the left  is a differential operator acting on the quantity to the right.  

The construction proceeds by looking at Hamiltonian evolution with respect to the Killing vector $\xi^a$.  The lapse and shift variables $F$  and $F^a$ are then given by the projections of the Killing vector perpendicular to and along the hypersurface, so that $F = -\xi^a n_a$ and $F^a  = \xi^a - Fn^a$.  The variation in the Hamiltonian is then $\delta H = F\delta H_\perp +F^c\delta H_c$ and  one can write
\begin{equation}\label{firstvariation}
\delta H 
 =  \left( {\delta H^*_\perp\over \delta s_{ab} }\cdot F
+ {\delta H^*_c\over \delta s_{ab}}\cdot F^c\right) \delta s_{ab}
+ \left( {\delta H^*_\perp\over \delta \pi^{ab} }\cdot F
+ {\delta H^*_c\over \delta\pi^{ab}}\cdot F^c\right)\delta\pi^{ab} +  D_c B^c,
\end{equation}
Here ${}^*$  denotes the adjoint differential operators that result from integrating by parts, and $ D_a$ is the covariant derivative operator compatible with the unperturbed spatial metric $\szero$.  

It follows from the Hamiltonian equations of motion that the quantities in parenthesis in (\ref{firstvariation}) are respectively $-\call_\xi \pi^{ab}_{(0)}$ and $\call_\xi \szero$, which vanish since
$\xi ^a$ is a Killing vector of the background. For vacuum spacetimes one has both $H=H_a =0$ and 
$\delta H=\delta H_a =0$ and therefore equation (\ref{firstvariation}) implies a Gauss-type law for
linearized solutions, namely
\begin{equation}\label{gauss}
D_c B^c =0. 
\end{equation}
Note that this applies in vacuum, while more generally perturbations to the stress energy tensor will appear on the right hand side.
Also note that the vector $B^a$ receives a contribution from each higher curvature term so that
 $ B^a = \sum_k b_k  B^{(k)a}$.
One finds in particular that  $B^{(0)a}=0$, while
\begin{equation}\label{bone}
B^{(1)a} = F(D^a h -D_b h^{ab} ) -hD^a F +h^{ab}D_b F + {1\over \sqrt{\bar s}}\left(F^a\pi^{bc}h_{bc} -\pi^{ab}h_{bc}F^c -2p^{ab}F_b\right) 
\end{equation}
gives the contribution from the Einstein term in the Lagrangian  \cite{Traschen:1984bp}.
Here we have set $h_{ab}=\delta s_{ab}$ and $p^{ab} = \delta\pi^{ab}$.  An explicit expression for the $B^{(k)a}$ for all $k$ is given in  \cite{Jacobson:1993xs}, but will not be needed here.
 
 The Gauss' law statement (\ref{gauss}) was used in reference \cite{Jacobson:1993xs} to derive the first law 
 in Lovelock theories with fixed $b_k $ and $G$ 
by integrating over the spatial surface.   As in the derivation of the first law in Einstein gravity \cite{Sudarsky:1992ty}, the boundary term at infinity is found to be proportional to the variation in the ADM mass, while the boundary term at the horizon is proportional to the surface gravity times the variation in the entropy.   Setting the whole integral equal to zero then gives the result for the first law with fixed Lovelock couplings in equation  (\ref{lovelockfirst}).

In order to derive the Smarr formula for Lovelock gravity, we now need to compute the contributions to the first law when the Lovelock coupling constants are allowed to vary.  We will consider independent variations in all the constants $b_k$ as well as in Newton's constant $G$.   Considering variation in the dimensionless coupling $b_1$ is redundant.  However, it makes the computations more uniform to carry through and we will see that the variation $\delta b_1$ drops out form the final result.

\subsection{Varying Newton's constant}

In order to understand how variations in Newton's constant enter the result, let us first reconsider the first law in
of Einstein gravity coupled to matter allowing for a non-zero $\delta G$.  Take the background
to be Minkowski spacetime and include the matter via a perturbative source term
 $\delta\rho$ (the background stress energy and perturbative current $\delta J_a$ are taken to  vanish). Adding the contributions of stress-energy into the perturbation theory presented above,  the Hamiltonian constraint equation $H=-16\pi G\rho$ linearizes to $D_a B^a = -16\pi \delta (G\rho )$,
  where $B^a =B^{(1)a} $ is given in equation (\ref{bone}). Integrating this result over a spatial slice, we have 
  \begin{equation}\label{easymass}
  \int _\infty da_c B^{(1)c} = -16\pi \delta(G M ).
  %-16\pi \int _V  dv \delta(G\rho ).
  \end{equation}
  Note that the background is Minkowski spacetime, so there is no internal boundary at a black hole horizon.
However, this result for the boundary term at infinity will continue to hold with a black hole present and therefore with $\delta G\neq 0$, the first law becomes
  \begin{equation}\label{easyfirst}
  \delta (GM)
  %\equiv  -{1\over 16\pi }  \int _\infty da_c B^c
   = {\kappa \over 8\pi} \delta A.
  \end{equation}
  Of course if $G$ is fixed then this is much ado about nothing. However, to derive
 the Smarr relation from the resulting first law, one must understand that the actual
 result is (\ref{easyfirst}). One gets the right result because the product $GM$ has dimension
  $L^{D-3}$  which is not true for the mass $M$ by itself.
  
  Applying this same reasoning to the results of \cite{Jacobson:1993xs} in Lovelock gravity, we 
 modify the equation for $\delta M$
 to replace $G\delta M$ with $\delta (GM)$ giving
  \begin{equation}\label{deltamfixedb}
  \delta ( GM ) = -{1\over 16\pi}
 \int _\infty da_c \sum _k  b_k  B^{(k)c} 
 \end{equation} 
 which holds for fixed couplings $b_k$.
 
\subsection{Varying Lovelock couplings}
Now consider perturbing the coefficients $b_k$ of the Lovelock theory as well.  The perturbed metric is assumed to satisfy the Lovelock equations of motion with the perturbed coefficients $b_k + \delta b_k$,
$G+\delta G$. For vacuum solutions one has $H=\delta H =0$.  Equation (\ref{gauss}) will receive a new contribution from the $\delta b_k$, giving
 \begin{equation}\label{newgauss}
  D_a B^a  + \sum_k \delta b_k (F \hk_\perp + F^a \hk _a ) =0
 \end{equation}
where $\hk_\perp$ and $\hk _a$ are evaluated with respect to the unperturbed variables $\bar s_{ab}$ and $\bar\pi^{ab}$.
Unlike $H_\perp$ and $H_a$ themselves, the sum above  with coefficients $\delta b_k$ rather than $b_k$  will not vanish in general.  This would appear to spoil the chances of deriving an expanded version of the first law from (\ref{newgauss}).  However, each term in the sum may be rewritten as a boundary term by making use of the Killing-Lovelock potentials $\beta^{(k)ab}$ defined in
 equation (\ref{potdef}).  
The result of integrating (\ref{newgauss}) over the spatial slice is then given by
\begin{equation}\label{gaussint}
 \int _{\partial V} da_c  \sum _k\left ( b_k B^{(k)c}[h_{ab}] + \delta b_k \beta^{(k)cd}n_d\right  ) =0.
\end{equation}
%
%where $h_{ab} = s_{ab} - s^{(0)} _{ab}$.
This result will now enable us to derive the generalized first law for Lovelock black holes including variations in the Lovelock couplings.  However, there are two issues that need to be resolved in order to process equation (\ref{gaussint}) into the first law.  
First is the issue of divergences.  With asymptotically AdS boundary conditions, each of the individual terms in (\ref{gaussint}) diverges at infinity.   In order to begin the process of interpreting the contributions at infinity, we will want to work with finite quantities.
Second, in Lovelock gravity both the mass (discussed below) and the entropy (\ref{entropy}) have explicit dependence on the Lovelock couplings that must be taken into account in identifying the variations $\delta M$ and $\delta S$.

\subsection{Dealing with divergences} \label{divergences}
When  the Lovelock couplings $b_k$  are varied, the curvatures of the Lovelock vacua discussed in section (\ref{vacua}) will vary as well.  
This leads to divergences in individual terms in the boundary term at infinity contained in equation (\ref{gaussint}).  In this section, we show that adding zero to the boundary integrand at infinity in a judicious way, yields finite quantities which will aid in the processing of the first law.  This procedure is described in some detail for the case  $\hat k=1$, {\it i.e.} Einstein gravity with $\Lambda<0$, in reference \cite{Kastor:2009wy}.  We will just sketch how it works out in the general Lovelock case.
%However by grouping the terms in (\ref{gaussint}) appropriately, these divergences may be shown to cancel.  

Let us assume that with Lovelock coupling 
constants $b_k$ the theory has an AdS vacuum with radius of curvature $l$.  
In static coordinates, the AdS vacuum metric is then given by
\begin{equation}\label{adsmetric}
ds^2 = - (1+ r^2/l^2) dt^2 + {dr^2\over (1+r^2/l^2)} + r^2 d\Omega_{D-2}^2.
\end{equation}
We will call the spatial part of this metric $\bar s^{A}_{ab}$.
Now consider varying the Lovelock couplings to nearby values $b_k + \delta b_k$.  The AdS curvature radius will generically also shift to a nearby value $l+\delta l$, where to good approximation $\delta l$ can be taken to depend linearly on the shifts $\delta b_k$ in the Lovelock couplings. 
We will call this new AdS vacuum metric $\bar s^{B}_{ab}$ and denote the difference between the two AdS metrics by $H_{ab} = \bar{s}^B_{ab} -\bar{s}^A_{ab}$.

Now consider the Gauss' law statement (\ref{gaussint}) for this setup.  Since their is no black hole horizon, the only contribution to the boundary integral comes from infinity and (\ref{gaussint}) implies that this integral at infinity must vanish.  Denoting the Killing-Lovelock potentials for the metric $\bar{s}^A_{ab}$ by $\beta _{AdS} ^{(k)ab}$, we then have the relation
\begin{equation}\label{adsinfinity}
 \int _\infty   da_c  \sum_k \left(  b_k B^{(k)c} (H_{ab}) + \delta b_k\, \beta _{AdS} ^{(k)cd} n_d \right) = 0.
\end{equation}
We have checked this formula explicitly by computing all the ingredients, each of which diverge in the limit that the boundary is taken to infinity, with the sum vanishing pointwise.  The check for the case $\hat k=1$ is carried out explicitly in \cite{Kastor:2009wy}.

%The perturbation to the metric (\ref{adsmetric}) will diverge as $r\rightarrow\infty$,
%and since the area element grows like $r^{D-2}$ the integral over the boundary diverges.
%  Denote the perturbation to the spatial metric in this case by $H_{ab} = \bar{s}^B -\bar{s}^A $ and denote the Killing-Lovelock potentials for the AdS vacuum by $\beta^{(k)ab}_{AdS}$.
%Since there is no black hole horizon in this case, equation (\ref{gaussint}) implies that
%% 
%\begin{equation}\label{adsinfinity}
% \int _\infty   da_c  \sum_k \left(  b_k B^{(k)c} (H_{ab}) + \delta b_k\, \beta _{AdS} ^{(k)cd} n_d \right) = 0.
%\end{equation}
%%
%Hence, divergences in the boundary terms $B^{(k)ab}$ at infinity for this perturbation to the spatial metric cancel precisely with the divergences in the Killing-Lovelock potentials of the AdS vacuum.

Now consider a static black hole with metric $s^A_{ab}$ in the Lovelock theory with couplings $b_k$ that is asymptotic at infinity to the AdS vacuum $\bar{s}^A_{ab}$, and also a perturbed black hole with metric $s^B_{ab}$ in the Lovelock theory with perturbed couplings $b_k+\delta b_k$ that is asymptotic to the 
AdS vacuum $\bar s^B_{ab}$.  The metric perturbation $h_{ab} = s^B_{ab} -s^A_{ab}$ between the black hole spacetimes then asymptotes at infinity to $H_{ab}$.  The cancellation of infinities in equation (\ref{adsinfinity}) guarantees that the total boundary term at infinity in (\ref{gaussint}) will be finite.  However, in order to work with manifestly finite contributions, we can subtract zero in the form of equation (\ref{adsinfinity}) from equation (\ref{gaussint}), giving
\begin{equation}\label{newgaussint}
0 = {1\over 16\pi G}\sum_k\left\{\int_\infty  da_c  b_k\left( B^{(k)c}[h_{ab}] - B^{(k)c}[H_{ab}]\right) - \int_H da_c b_k B^{(k)c}[h_{ab}] -   \delta b_k\Theta^{(k)}\right\}
\end{equation}
where we have grouped together the Killing-Lovelock potential terms into the quantities
\begin{equation}\label{thetak}
\Theta ^{(k)} = \int _H da_c \beta^{(k)cd} n_d  - \int _\infty  da_c  (\beta^{(k)cd} - \beta _{AdS} ^{(k)cd} )n_d .
\end{equation}
%
%%
%\begin{eqnarray}\label{newgaussint}
%0 &=& {1\over 16\pi G}\int_\infty  da_c \sum_k \left\{ b_k\left( B^{(k)c}[h_{ab}] - B^{(k)c}[H_{ab}]\right) + \delta b_k\left (\beta^{(k)cd}-\beta^{(k)cd}_{AdS}\right) \right\}\\
%&& - {1\over 16\pi G} \int_H da_c\sum_k\left\{ b_k B^{(k)c}[h_{ab}] + \delta b_k\beta^{(k)cd} \right\}.\nonumber
%\end{eqnarray}
%%
The terms grouped by parenthesis in the integrals at infinity in (\ref{newgaussint}) and (\ref{thetak}) can be shown to sum to finite quantities.  
%The quantities $\Theta^{(k)}$ are independent of the ambiguities in the choice of the Killing-Lovelock potentials noted in section (\ref{killingpotential}) and discussed in more detail in reference \cite{Kastor:2008xb}.
With the Gauss' law identity processed into the form (\ref{newgaussint}), we are now in a position to identify the contributions to the variations in the mass and entropy and establish the extended first law.

%Now let the background spacetime be a static AdS black hole and
% consider perturbations amongst asymptotically AdS black holes in which the Lovelock couplings $b_k$ vary.  As above, the background and perturbed black hole solutions will approach AdS vacua with slightly different curvatures.  Asymptotically, the perturbation to the spatial metric will approach the quantity $H_{ab}$  defined above and this will lead to divergences in the integrals of the boundary terms $B^{(k)a}$ at infinity.  These divergences, however, are cancelled as in equation (\ref{adsinfinity}) by grouping them together with the divergent terms in the Killing-Lovelock potentials.  Accordingly, we divide the Killing-Lovelock potentials at infinity into two pieces, $\beta^{(k)ab}_{AdS}$ and $\beta^{(k)ab}-\beta^{(k)ab}_{AdS}$. The first piece combines with the boundary term $B^{(k)a}$ to yield a finite boundary integral at infinity.  This is important in identifying $\delta M$ in the perturbative Gauss's law identity (\ref{gaussint}),
%which we turn to now.

\subsection{Extended first law and Smarr formula}

The task at hand is now to relate the Gauss' law expression (\ref{newgaussint}) to the variations in the mass and entropy.
In order to understand the variation $\delta M$ in the mass for Lovelock black holes, we need to start with an expression for the mass $M$ itself.
As discussed in reference  \cite{Jacobson:1993xs},
%To process the Gauss' law identity (\ref{gaussint}) into the first law one needs
%to know what $\delta M$ is. For Einstein gravity with a cosmological constant
%this is well known, and in particular the cosmological constant does not enter explicitly into
%the expression for the ADM mass. However in a more general Lovelock  AdS  theory the
%parameters $b_k , k\geq 2$ do appear explicitly. Therefore we need to derive 
%the general  mass formula in order to interpret our preliminary form of the first law.
the Hamiltonian formulation of Lovelock gravity is well-suited to deriving an expression for
the mass. Following the reasoning of  Regge and Teiltelboim \cite{Regge:1974zd} in Einstein gravity, in order to have a well-defined variational principle, the volume term of the Lovelock Hamiltonian must be supplemented
by a boundary term, whose variation cancels the boundary term that arises in the variation of the bulk Hamiltonian.  The mass $M$ of the spacetime is given by the value of the Hamiltonian.  Since the bulk Hamiltonian vanishes on solutions, the mass is given solely by the boundary term.    
%so that the variation of the total Hamiltonian is well defined.
The steps required to obtain the required Hamiltonian boundary term are essentially those of 
%one writes the variation of  $\int H$ as the adjoint operator plus
%a total derivative, which then becomes a boundary term. But these are just the steps
the Hamiltonian perturbation theory that produces $B^a$ in (\ref{firstvariation}).
For a metric $s^A_{ab}$, that is asymptotic at infinity to the AdS vacuum with metric 
$\bar s^A_{ab}$, the mass  \cite{Jacobson:1993xs} is then given by 
\begin{equation}\label{admmass}
M [s^A_{ab} ] = -{1\over 16\pi G}\sum_k b_k
\int _\infty da_c B^{(k)c} [s^A_{ab}-\bar{s}^A_{ab}].
\end{equation}
Note that the explicit dependence of the mass on the Lovelock couplings will contribute to $\delta M$ when these couplings are varied.

Now assume that the nearby metrics $s^A_{ab}$ and $s^B_{ab}$ are as described in section (\ref{divergences}).  Recalling that the boundary vectors $B^{(k)a}$ are linear in their arguments, the variation in the mass $\delta M= M[s^B_{ab}] - M[s^A_{ab}]$  is  given by 
\begin{eqnarray}
\delta M &=& - {1\over 16\pi G} \sum_k\int_\infty da_c\left\{ b_k B^{(k)c}[(s^B_{ab}-\bar{s}^B_{ab})-(s^A_{ab}-\bar{s}^A_{ab})]
+\delta b_k B^{(k)c}[s^A_{ab}-\bar{s}^A_{ab}]\right\} \\
&=& -{1\over 16\pi G} \sum_k\int_\infty da_c\left\{ b_k \left(B^{(k)c}[h_{ab}]-B^{(k)c}[H_{ab}]\right)
+\delta b_k B^{(k)c}[s^A_{ab}-\bar{s}^A_{ab}]\right\}.\label{deltam}
\end{eqnarray}

The other ingredient we need is the variation of  the expression (\ref{entropy}) for the entropy of a Lovelock black hole, which is given by
$\delta S = {1\over 4 G}\sum_k\left( b_k\delta A^{(k)} +A^{(k)}\delta b_k\right).$
%%
%\begin{equation}
%\delta S = {1\over 4 G}\sum_k\left( b_k\delta A^{(k)} +A^{(k)}\delta b_k\right)
%\end{equation}
%%
From the derivation of the first law with fixed Lovelock couplings  \cite{Jacobson:1993xs}, we know that 
%$\delta A^{(k)} = -{1\over 2\kappa} \int_H da_c B^{(k)c}[s^B_{ab} - s^A_{ab}]$
%
\begin{equation}
\delta A^{(k)} = -{1\over 2\kappa} \int_H da_c B^{(k)c}[s^B_{ab} - s^A_{ab}]
\end{equation}
while the black hole temperature is related to the horizon surface gravity according to $T=\kappa/2\pi$.  Therefore we have
\begin{equation}\label{tds}
T\delta S = -{1\over 16\pi G} \int_H da_c \sum_k b_kB^{(k)c}[s^B_{ab} - s^A_{ab}] +{\kappa\over 8\pi G}\sum_k A^{(k)}\delta b_k
\end{equation}

We can now substitute the expressions (\ref{deltam}) and (\ref{tds}) for $\delta M$ and $T\delta S$ into the Gauss' law relation (\ref{newgaussint}) to obtain a first law 
having the form given in equation (\ref{firstlawform}).
%%
%\begin{equation}
%\delta M = T\delta S - {1\over 16\pi G}\sum_k \Psi^{(k)}\delta b_k
%\end{equation}
%%
If we make the further definition 
\begin{equation}
B^{(k)} = \int_\infty da_c B^{(k)c}[s^A_{ab}-\bar s^A_{ab}] 
\end{equation}
then the thermodynamic potentials $\Psi^{(k)}$ are given by
\begin{equation}\label{psik}
\Psi^{(k)} = 2\kappa A^{(k)} + B^{(k)}+ \Theta^{(k)} ,
\end{equation}
%
%and the $\Theta^{(k)}$ are in turn given in terms of the Killing-Lovelock potentials by
%%
%\begin{equation}\label{theta}
%\Theta ^{(k)} = \int _{\hor}  da r_c \beta^{(k)cd} n_d  - \int _\infty  da r_c  (\beta^{(k)cd} - \beta _{AdS} ^{(k)cd} )n_d .
%\end{equation}
%
with the three terms originating respectively from the dependence of  the entropy, the mass and the bulk Hamiltonian on the 
Lovelock couplings.  Note that the quantities $B^{(0)c}$ and $A^{(0)}$ both vanish and that consequently this general result reduces to that of reference \cite{Kastor:2009wy} in the case of Einstein gravity with $\Lambda<0$.  The full Smarr formula can now be assembled from (\ref{lovelocksmarr}) and (\ref{psik}) and has the form
\begin{equation}\label{finalsmarr}
(D-3) M = (D-2) TS - \sum_k {2(k-1)\over 16\pi G}\left(  2\kappa A^{(k)} + B^{(k)}+ \Theta^{(k)}\right) b_k
\end{equation}

The formula (\ref{psik}) gives  geometrical expressions for the potentials $\Psi^{(k)}$ that are thermodynamically conjugate to the Lovelock couplings $b_k$.  One obvious comment is that these expressions are more complicated then the thermodynamic quantity conjugate to the entropy ({\it i.e.} the temperture) which remains simple in Lovelock gravity.  It seems possible that there exists a simpler formulation for the $\Psi^{(k)}$.  In this context, it is worth noting that the quantities $B^{(k)}$ can each be shown to be proportional to the mass \cite{Kastor:inprep} according to
\begin{equation}
B^{(k)} = - k {16\pi G (D-1)!\over  \sigma_1\, (D-2k-1)! } \left( {-1\over l^2}\right )^{k-1} M
\end{equation}
where $\sigma_1 =  \sum_k {(D-1)! \over(D-2k-1)! } \left( {-1\over l^2}\right )^{k-1}k\, b_k $. 

\subsection{Physical significance of the potentials $\Theta^{(k)}$}\label{volpots}

The terms $A^{(k)}$ and $B^{(k)}$ in the expression (\ref{psik}) for $\Psi^{(k)}$ are somewhat familiar, being related to the entropy and mass respectively.  
The terms $\Theta^{(k)}$, on the other hand, are new and we should ask
what intuition can we gain into them? 
%new terms proportional to the potentials $\Theta^{(k)}$ that appear in the extended first law (\ref{firstlawform}) and Smarr formula (\ref{smarr})?  
Let us start by drawing attention to two important features of the formula (\ref{thetak}) for the $\Theta^{(k)}$'s.  

First, recall that the anti-symmetric Killing-Lovelock potentials $\beta^{(k)ab}$ are  specified only up to the addition of an arbitrary divergenceless anti-symmetric tensor.  The potentials 
$\Theta^{(k)}$, however, are invariant under such shifts (see the discussion of this point in reference \cite{Kastor:2009wy}).  
Secondly, the integrals of $\beta^{(k)ab}$ and $\beta^{(k)ab}_{AdS}$ at infinity are individually divergent.  To obtain a finite result, one must take the difference of the two integrals over finite sized spheres and then take the limit in which the radius of the sphere becomes infinite.  The potentials $\Theta^{(k)}$ then represent `renormalized' or `effective' quantities.

The physical significance of the potentials $\Theta^{(k)}$ is perhaps easier to assess by converting the boundary integrals in (\ref{thetak}) back into volume integrals.
 Making use of equation (\ref{kthhamdiv}), one can express the $\Theta^{(k)}$ in terms of volume integrals of the individual terms $H^{(k)}$ in the Hamiltonian. 
The $k^{th}$ Hamiltonian function is given explicitly by
\begin{equation}
H^{(k)} = -{1\over 2^k } \tilde{  \delta } ^{a_1 b_1\dots a_k b_k } _{c_1 d_1 \dots c_k d_k }\,
 R_{a_1 b_1}{}^{c_1 d_1 } [g]  \dots  R_{a_k b_k}{}^{c_k d_k } [g].
 \end{equation}
where the tilde denotes the anti-symmetrized delta-function projected with the spatial metric $ s^a _b $ on all its indices.
The Riemann tensors here are those of the full spacetime metric. However, because
all indices on the Riemann tensors are projected with the spatial metric, we can use the Gauss-Codazzi equations to write
them in terms of the Riemann tensor for the spatial metric $s_{ab}$ together with the extrinsic curvature $K_{ab}$
of the spatial slice according to $\tilde{R}_{ab}{}^{cd} [g]= R_{ab}{}^{cd} [s] + K_{[a} {}^c K_{b]}{} ^d $.
Now recall that the spacetime metric is assumed to have a static Killing vector $\xi^a$.  Let us choose a spatial slice $\Sigma$ with normal
in the direction of the Killing field, so that $\xi ^a =F n^a$. For this slice the extrinsic curvature
$K_{ab}$ vanishes and the $k^{th}$ Hamiltonian becomes
\begin{equation}\label{specialhk}
H^{(k)}  = -{1\over 2^k } \tilde{  \delta } ^{a_1 b_1\dots a_k b_k } _{c_1 d_1 \dots c_k d_k }\,
 R_{a_1 b_1}{}^{c_1 d_1 } [s]  \dots  R_{a_k b_k}{}^{c_k d_k } [s]. 
 %\equiv -L_k [s]
\end{equation}
This is the same as minus the $k^{th}$ Lovelock Lagrangian  given in (\ref{lovelagran}), but now evaluated on the spatial metric ({\it i.e.} for this configuration $H^{k}= - \callk$).   Combining equations (\ref{kthhamdiv})  and (\ref{specialhk}) 
the formula  (\ref{thetak}) for the $\Theta^{(k)}$ can now be written as
\begin{equation}\label{thetavol}
\Theta^{(k)} =\int _\Sigma \sqrt{-g} \callk [s] - 
\int _{\Sigma_{AdS}} \sqrt{-g _{AdS}} \callk [s_{AdS}]
\end{equation}
In the first integral, the spatial slice $\Sigma$  runs from the black hole horizon out to spatial infinity, while in the second integral $\Sigma_{AdS}$ is a spatial slice of AdS and has no internal boundary.
In both integrals the slice is chosen to have unit normal such that $\xi ^a =F n^a $,
and we have combined $\sqrt{-g} =F\sqrt{s}$. As noted above, each of the volume integrals in (\ref{thetavol}) is divergent and therefore the formula must be understood as integrating out to some large spheres in the asymptotic region and carrying out the subtraction before letting
the spheres go to infinity.

The potential $\Theta^{(0)}$ was previously considered in reference  \cite{Kastor:2009wy}. 
The $0^{th}$ Lovelock Lagrangian is simply $\call^{(0)}=1$ and therefore the integrals in (\ref{thetavol}) are the volumes of the 
respective slices.  Since the slice $\Sigma_{AdS}$ covers the whole of the interior, while the slice $\Sigma$ has the interior of the black hole removed,
we can think heuristically of the difference in (\ref{thetavol}) as minus an effective volume $V_{eff}$ of the black hole interior,
 \begin{equation}\label{thetazero}
\Theta^{(0)} = -V_{eff}.
\end{equation}
As noted in \cite{Kastor:2009wy}, thinking about $\Theta^{(0)}$ in this way leads to a thermodynamic interpretation in the case of Einstein gravity with a nonzero cosmological constant $\Lambda=-b_0/2$.
The cosmological constant is naturally associated with a pressure through $p = - \Lambda/8\pi G$ and the new terms in the extended first law and 
Smarr formula in this case become respectively $V_{eff} \delta p$ and $p V_{eff}/4\pi G$.  The form of this contribution to the first law led us to the interpretation of the mass $M$ as a kind of enthalpy in the context of a variable cosmological constant.

 \section{Concluding Remarks and Future Directions}\label{secfree}
We have used geometric methods to establish certain properties of static, asymptotically AdS black holes in Lovelock gravity theories, 
namely a Smarr formula (\ref{finalsmarr}) and a related extension of the first law (\ref{firstlawform}) that includes variations in the Lovelock couplings.
In addition to the familiar term proportional to the surface gravity times the area, the Smarr formula\footnote{A Smarr formula for AdS black strings in Einstein-Gauss-Bonnet gravity was found using boundary counterterm methods  in \cite{Brihaye:2008xu}which does not include the $\Psi^{(k)}$-type potential terms found here.  
The solutions studied in \cite{Brihaye:2008xu}, including a compact direction, differ in boundary conditions from those considered here and hence our results cannot be compared directly.  However, we find the absence of the potential terms in the results of \cite{Brihaye:2008xu} surprising.  We expect this is due to the use of different definitions of the mass.} contains a sum over a set of geometric potentials $\Psi^{(k)}$ 
%that are $k^{th}$ order in curvature 
multiplied by the Lovelock couplings $b_k$.  A similar sum appears in the first law.

The new potentials $\Psi^{(k)}$ in (\ref{psik}) include three types of contributions which we have written as $A^{(k)}$, $B^{(k)}$ and $\Theta^{(k)}$, related to the entropy, mass and Killing-Lovelock potentials respectively.  The terms $\Theta^{(k)}$, in particular,  are expressed in terms of boundary integrals of the anti-symmetric Killing-Lovelock potentials $\beta^{(k)ab}$ defined in (\ref{potdef}).  We have shown that they can be re-expressed as in (\ref{thetavol}) as  finite, renormalized volume integrals of the Lovelock Lagrangian densities over a spatial slice extending from the horizon out to infinity.  The AdS/CFT correspondence offers a potentially rewarding direction to look for further insight into the physical significance of the $\Psi^{(k)}$'s.  It is well known that the properties of Einstein-AdS black holes are related to those of the high temperature phase of the boundary CFT \cite{Witten:1998qj,Witten:1998zw}.  As noted in the introduction, the impact of Gauss-Bonnet and higher derivative Lovelock terms on the boundary CFT has been the subject of substantial recent interest\footnotemark[2].   The potentials $\Psi^{(k)}$ should be relevant to this story.

A related application of  the Smarr formula (\ref{finalsmarr})  that we plan to explore in future work  is using it as an aid in the analysis of the free energy of Lovelock black holes.  
Hawking and Page \cite{Hawking:1982dh}
showed that sufficiently large AdS-Schwarschild black holes have positive specific heat and are therefore thermodynamically  stable. Further, they 
computed the 
free energy  $F$ 
as a function of
the horizon radius $r_H$ and $\Lambda$,
finding a phase transition where the free energy for large black holes relative to the AdS background becomes negative above a certain temperature. 
Through the AdS/CFT correspondence in a $D=5$ bulk, this Hawking-Page phase transition signals the transition from the low temperature, confining phase to the high temperature, de-confining phase of the gauge theory on the boundary \cite{Witten:1998qj,Witten:1998zw}.
Similar analyses have been done for Gauss-Bonnet-AdS
black holes \cite{Cvetic:2001bk,Cai:2001dz,Nojiri:2001aj,Cho:2002hq} using the known analytic solution \cite{Boulware:1985wk,Wheeler:1985nh}\footnote{See also references \cite{Cai:1998vy,Cai:2003gr,Cai:2003kt} for further discussions of the thermodynamic and stability properties of Lovelock black holes}. The calculations are significantly more complicated and yield a more intricate set of results.  
Only  some parts of the $(b_0 , b_2 )$ Lovelock parameter space admit stable, large black holes.
For still higher curvature Lovelock theories, the static black hole solutions are known only implicitly in terms of a metric function that solves a higher order polynomial equation \cite{Wheeler:1985qd}.  In this case, it is both more challenging (or perhaps not possible) to compute the free energy explicitly and also potentially more challenging to analyze any results over the expanded Lovelock parameter space.

 However, the process can be simplified.
 The free energy is given by the temperature $T$ times
 the Euclidean action $I_E$. Working in the Hamiltonian picture one can show that
  $I_E =\beta M -S$ where $\beta$ is the Euclidean period  \cite{Hawking:1995fd}. 
For a smooth Euclidean horizon the period is equal to the inverse
temperature
and one has $F = M-(\kappa /2\pi) S$.
The Smarr formula  (\ref{finalsmarr})  can now be used to eliminate the mass from the expression of 
$F$, giving
\begin{equation}\label{free}
(D-3) F =TS- {1\over 4\pi G} \sum_k b_k
(k-1) \Psi^{(k)} .
\end{equation}
In the Hawking-Page case only the parameters $b_0$ and $b_1$ are
nonzero, and using the relation $b_0=-2\Lambda$ this reduces to
\begin{equation}\label{hpfree}
F =   {2\Omega _{D-2} \over D-3}  r_H ^{D-2} \left( \kappa +  {2\Lambda   r_H \over D-1 } \right) 
\end{equation}
In this form we see that there is a positive contribution to the free energy coming from the area, 
and a negative contribution (for $\Lambda<0$) coming from
the potential $ \Theta^{(0)}$. Which one of these terms dominates depends on how the surface
gravity $\kappa$ depends on the horizon radius.  We have therefore gained some geometrical understanding of the phase transition, namely that 
the possibility of a thermodynamically
stable  large black hole comes from the 
$ \Theta^{(0)}$ term in the Smarr relation. 
In the limits of large and small AdS black holes respectively, it turns
out that 
\begin{equation}\label{surfgravlimits}
\kappa \approx {D-3\over 2 r_H},  \quad b_0 r_H ^2 \ll 1 ,\qquad
\kappa \approx  {b_0 r_H \over 2( D-2) },  \quad b_0 r_H ^2 \gg 1
\end{equation}
Substituting these into (\ref{hpfree}) one finds that the free energy $F$ is indeed negative for
the large black holes, signaling the phase transition.

This  analysis suggests that equation (\ref{free}) might be used to study the properties of the free energy $F$
for a general Lovelock black hole, even in the absence of fully explicit, analytic solutions.
In order to analyze the thermodynamics,  one would need to know $F$
as a function of $r_H$ and the Lovelock couplings $b_k$. There are two obstacles to this plan.
The surface gravity and the potentials $\Psi^{(k)}$ both need to be determined as functions of $r_H$,
in the absence of analytic solutions. We will pursue this goal, and also its extension to 
rotating black holes, in future work.

\subsection*{Acknowledgements}

The authors would like to thank Julio Oliva, Steven Willison and Jorge
Zanelli for useful discussions.

The work of DK and JT  is supported in part by NSF grant PHY-0555304.   The work of SR was funded by FONDECYT grant 3095018 and by the CONICYT
grant \textquotedblleft Southern Theoretical Physics Laboratory
\textquotedblright\ ACT-91. Centro de Estudios Cient\'{\i}ficos (CECS)
is funded by the Chilean Government through the Millennium Science
Initiative and the Centers of Excellence Base Financing Program of
CONICYT. CECS is also supported by a group of private companies which at
present includes Antofagasta Minerals, Arauco, Empresas CMPC, Indura,
Naviera Ultragas and Telef\'{o}nica del Sur. CIN is funded by CONICYT
and the Gobierno Regional de Los R\'{\i}os.

\end{document}